\documentclass[conference,ano]{IEEEtran}
\usepackage{amsmath,amsfonts}
\usepackage{array}
\usepackage[caption=false,font=normalsize,labelfont=sf,textfont=sf]{subfig}
\usepackage{textcomp}
\usepackage{stfloats}
\usepackage{url}
\usepackage{verbatim}
\usepackage{graphicx}
\usepackage[colorlinks=true, linkcolor=black, citecolor=black, urlcolor=black]{hyperref}
\usepackage{amstext,amssymb,amsfonts,latexsym}
\usepackage{graphicx}
\usepackage{enumitem} 
\usepackage{array} 
\usepackage{caption} 
\usepackage{ccicons}
\usepackage{animate}
\usepackage{setspace}
\usepackage{longtable}
\usepackage{graphicx}
\usepackage{booktabs}
\usepackage{pifont}
\usepackage[utf8]{inputenc}
\usepackage{rotating}
\usepackage{tabularx}
\usepackage{multirow}
\usepackage{color}
\usepackage[normalem]{ulem}
\usepackage{tabularray}
\usepackage{amsmath}
\usepackage{cite}
\usepackage{float}
\usepackage{placeins}
\IEEEoverridecommandlockouts
\usepackage{cite}
\usepackage{amsmath,amssymb,amsfonts}
\usepackage{graphicx}
\usepackage{textcomp}
\usepackage{xcolor}
\usepackage{algorithm}
\usepackage{algpseudocode}
\def\BibTeX{{\rm B\kern-.05em{\sc i\kern-.025em b}\kern-.08em
    T\kern-.1667em\lower.7ex\hbox{E}\kern-.125emX}}
\begin{document}

\title{RGD: Multi-LLM Based Agent Debugger via Refinement and Generation Guidance}
\author{
\IEEEauthorblockN{1\textsuperscript{st} Haolin Jin}
\IEEEauthorblockA{\textit{University of Sydney} \\
hjin3177@uni.sydney.edu.au}
\and
\IEEEauthorblockN{2\textsuperscript{nd} Zechao Sun}
\IEEEauthorblockA{\textit{University of Sydney} \\
zsun6058@uni.sydney.edu.au}
\and
\IEEEauthorblockN{1\textsuperscript{*} Huaming Chen}
\IEEEauthorblockA{\textit{University of Sydney} \\
huaming.chen@sydney.edu.au}
}
\maketitle
\begin{abstract}
Large Language Models (LLMs) have shown incredible potential in code generation tasks, and recent research in prompt engineering have enhanced LLMs' understanding of textual information. However, ensuring the accuracy of generated code often requires extensive testing and validation by programmers. While LLMs can typically generate code based on task descriptions, their accuracy remains limited, especially for complex tasks that require a deeper understanding of both the problem statement and the code generation process. This limitation is primarily due to the LLMs’ need to simultaneously comprehend text and generate syntactically and semantically correct code, without having the capability to automatically refine the code. In real-world software development, programmers rarely produce flawless code in a single attempt based on the task description alone, they rely on iterative feedback and debugging to refine their programs. Inspired by this process, we introduce a novel architecture of LLM-based agents for code generation and automatic debugging: Refinement and Guidance Debugging (RGD). The RGD framework is a multi-LLM-based agent debugger that leverages three distinct LLM agents—Guide Agent, Debug Agent, and Feedback Agent. RGD decomposes the code generation task into multiple steps, ensuring a clearer workflow and enabling iterative code refinement based on self-reflection and feedback. Experimental results demonstrate that RGD exhibits remarkable code generation capabilities, achieving state-of-the-art performance with a 9.8\% improvement on the HumanEval dataset and a 16.2\% improvement on the MBPP dataset compared to the state-of-the-art approaches and traditional direct prompting approaches. We highlight the effectiveness of the RGD framework in enhancing LLMs' ability to generate and refine code autonomously.

\end{abstract}

\section{Introduction \label{cha:intro}}
\begin{figure*}[htbp]
\footnotesize
    \centering
    \includegraphics[width=.8\textwidth]{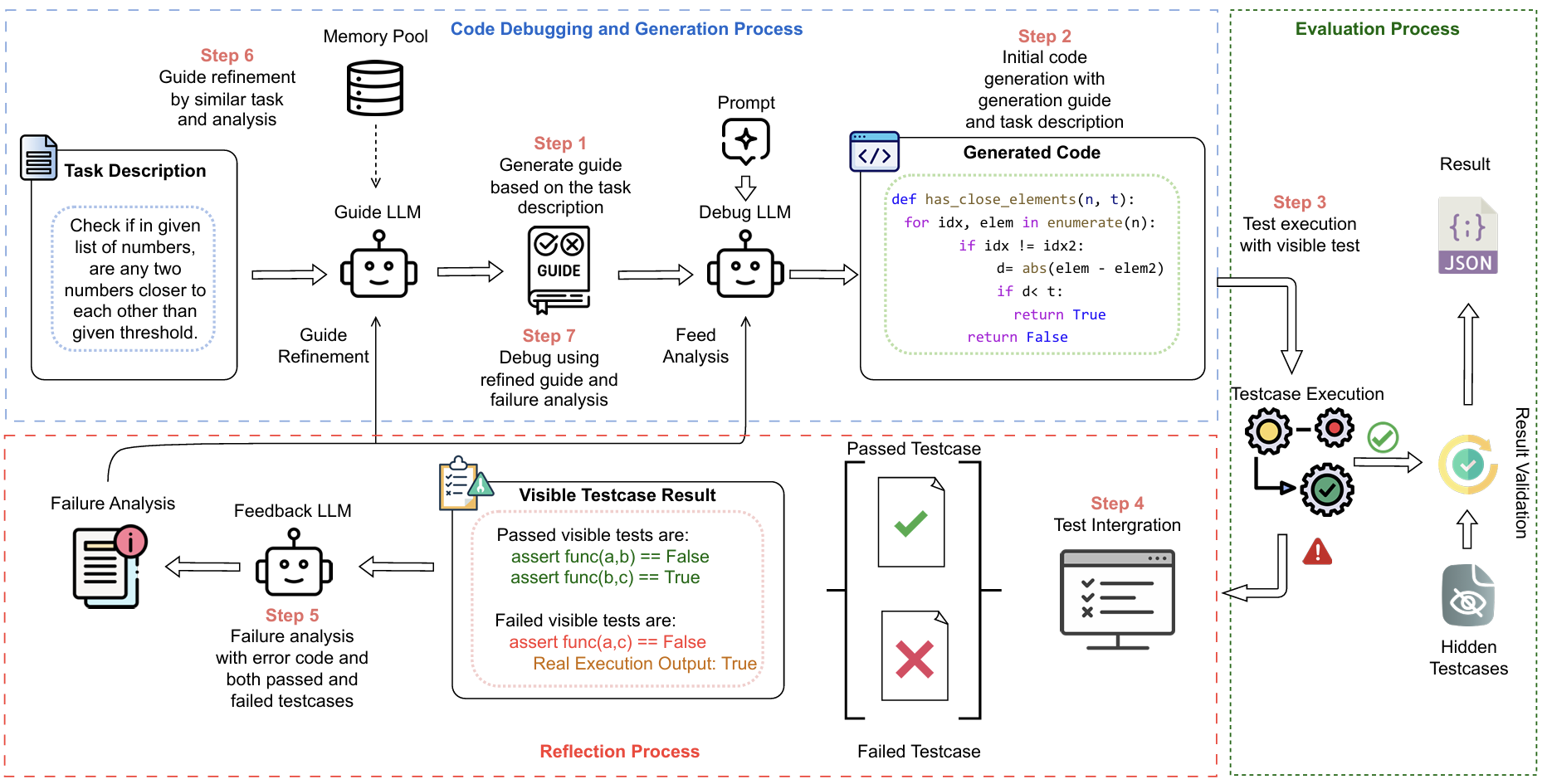} 
    \caption{The overview of the RGD framework contains three processes: the code generation and debugging process, the evaluation process, and the reflection process. The Guide LLM is responsible for both generating guides and retrieving relevant information from the memory pool. The Debug LLM uses this information to generate code and incorporate failure analysis from the Feedback LLM to fix the code. The generated code is tested against both visible and hidden test cases to ensure comprehensive coverage and accuracy. This process is iterated until the code passes all visible and hidden test cases or reaches the maximum number of iterations.} 
    \label{fig:framework}
\end{figure*}
Large Language Models (LLMs) have made significant advancements in the domain of automated code generation, showcasing their capability to translate natural language into functional code, generate code explanations \cite{ni2023l2cevalevaluatinglanguagetocodegeneration}, and even perform code-to-code translations across different programming languages \cite{sun2023sql} \cite{zheng2024codegeexpretrainedmodelcode}. The most prevalent methods for employing LLMs in code generation rely heavily on prompt engineering, where well-crafted prompts are designed to guide the LLM in generating code snippets or interpreting existing code \cite{chen2021evaluating} \cite{wei2022chain}. This approach has proven effective in a range of scenarios, from generating code based on textual descriptions to converting code between different languages and frameworks.

However, the approach of generating code from text in a single pass has its limitations. Code generation is inherently a complex task, and a one-time generation approach often fails to account for the numerous edge cases and detailed task requirements that arise in software development, especially given the complexity and precision needed in software development tasks \cite{10.1145/3672456}. Consequently, researchers have introduced multi-round code generation frameworks that involve iterative refinement. These frameworks iteratively generate programs through multiple interactions, significantly improving the quality of program synthesis and making the development process more efficient and accurate \cite{nijkamp2023codegenopenlargelanguage}. Researchers have explored ways for LLMs to autonomously learn from errors and perform debugging and repairs. These frameworks leverage reflection, including failed test cases and error messages, and learns from these outcomes to improve subsequent code generations \cite{chen2023teaching}. Although these approaches have demonstrated significant improvements, they cannot guarantee that every reflection result will lead the LLM to make effective changes based on failed test cases. As a result, LLMs often end up generating the same code over multiple iterations.

Moreover, the performance of LLMs in code generation is highly dependent on the clarity and completeness of the task description provided in the prompts, there is a high likelihood of the LLM overlooking critical edge cases or missing essential requirements. Previous studies have attempted to address this issue by introducing the concept where the LLM first reasoning the task and then proceeds with code generation \cite{yao2022react} \cite{10.1145/3672456}. However, this strategy presents its own challenges; the LLM tends to rely heavily on the initial plan, leading to a lack of flexibility as it fails to incorporate further refinements or adjustments that may be necessary to address evolving requirements or unforeseen issues in the code.

In this paper, we introduce a novel framework called RGD (Refinement and Guidance Debugging) that leverages multiple LLMs in a collaborative manner to improve the quality of code generation. RGD incorporates multiple specialized LLMs, each with distinct roles, to simulate a comprehensive code repair process. The RGD framework utilizes three LLMs: the Guide LLM, responsible for generating a generation guide based on the task description, then passed the generation guide and the task description to the Debug LLM for initial code generation. The third LLM is tasked with consolidating the execution results and conducting failure analysis. The feedback LLM will use generated code and failed test cases to analysis the reason of failure and potential fix procedures. In addition to having the LLM analyze the reasons for failed test cases, our architecture also considers the test cases that pass. This step ensures that while the LLM analyzes the failed test cases, it is also aware of how the code passes the correct test cases, thereby preventing future code generations from failing previously passing test cases. This situation is very common, if there are only failed test sets and executed code, LLM will focus on trying to make the code successfully match the expected output and may not be able to actually find the problem in the code to debug, so the changed code will fail all test sets, and the failure analysis is also to avoid this scenario.

Furthermore, we draw inspiration from the RAG (Retrieval-Augmented Generation) approach, where we match similar content from a memory pool as additional information \cite{lewis2020retrieval}. Unlike traditional applications where the retrieved content is often code, we store relevant generation guides and task descriptions. This allows the Guide LLM to generate guidance that avoids zero-shot learning scenarios and instead relies on past successful examples to produce high-quality responses. Figure \ref{fig:framework} shows the overall flow and RGD architecture.

To evaluate the RGD framework, we conducted tests on various benchmarks, including HumanEval \cite{chen2021evaluating}, HumanEval-ET \cite{dong2023codescore}, MBPP \cite{austin2021programsynthesislargelanguage}, MBPP-ET \cite{dong2023codescore}, and APPS \cite{hendrycksapps2021}. These datasets are all text-to-code generation tasks and contain tasks and test sets of varying difficulty levels. HumanEval-ET and MBPP-ET are extended versions with additional edge case test cases to address the limitations of the original test sets. Through our experiments, we evaluated the performance of GPT-4o and GPT-4o-mini under the RGD framework across multiple datasets. The results demonstrated significant improvements of RGD on all benchmarks, surpassing the current state-of-the-art frameworks including LDB \cite{zhong2024ldb}.

\section{Related Work \label{cha:related}}
\textbf{Large Language Models for Code Generation} Recently, large language models (LLMs) have demonstrated outstanding performance multiple tasks especially in the code generation, showing great potential across various benchmarks \cite{chen2021evaluating} \cite{roziere2023code} \cite{yeticstiren2023evaluating} \cite{jin2024llmsllmbasedagentssoftware}. However, when faced with complex problems, these models are prone to generating hallucinated responses \cite{10.1145/3649506}. Prompt engineering \cite{white2023prompt} has reduced the need for extensive fine-tuning of LLMs on specific downstream tasks, which can better understand user requirements through contextual information. Chain-of-thought prompting \cite{10.1145/3672456}, which guides LLMs to provide step-by-step responses, enhances their reasoning abilities, leading to significant improvements over baselines on multiple benchmarks \cite{huang2024codecottacklingcodesyntax}. Tree-of-thought (ToT) \cite{yao2024tree} framework enhances the reasoning capabilities of large language models by structuring the reasoning process as a tree search. Each node represents a possible reasoning state, and edges denote transitions between states. In addition, researchers have explored enabling LLMs to debug the code they generate on their own \cite{hu2024leveragingprintdebuggingimprove}, allowing the models to autonomously fix errors by understanding the causes from generated outputs \cite{10.1145/3649825} \cite{chen2023teaching}. Beyond self-repair based on execution results, these frameworks can also leverage external tools for dynamic and automatic code correction \cite{wu2023autogen} \cite{olausson2023self} \cite{jiang2023selfevolve}.

\textbf{Information Retrieval} is a popular strategy recently, which involves retrieving helpful information from local storage or the cloud to assist in generation tasks \cite{he2024conline}. It has gained significant traction in LLM-based agents. For example, RAG \cite{gao2023retrieval} \cite{guu2020retrieval} and Microsoft's GraphRAG \cite{procko2024graph} transform the original zero-shot learning generation environment into a few-shot learning scenario. Additionally, the matching algorithms can be adjusted to better fit specific task requirements. By leveraging web queries and utilizing tools to collect data in real-time from the internet \cite{zhang2023toolcoder}, the capability of LLMs is further enhanced. RAG has shown remarkable performance in downstream tasks like code generation \cite{fan2024survey}. By ranking and sort the retrieved information, it increases reliability and reduces the impact of hallucinations \cite{islam2024mapcoder}.
\section{Refinment and Guidance Debugging\label{cha:framework}}
\subsection{Definitions}
Following recent works \cite{zhong2024ldb} \cite{chen2023teaching}, we divide the benchmarking dataset HumanEval and MBPP into three components: $(Q, T_v, T_h, E)$. Here, $Q$ represents the task description, which includes code snippets and requirements in natural language. $T_v$ stands for visible test cases, $T_h$ for hidden test cases, and $E$ is the entry point. All test cases are executed starting from the entry point.
\begin{figure}[htbp]
\scriptsize
    \centering
    \includegraphics[width=.4\textwidth]{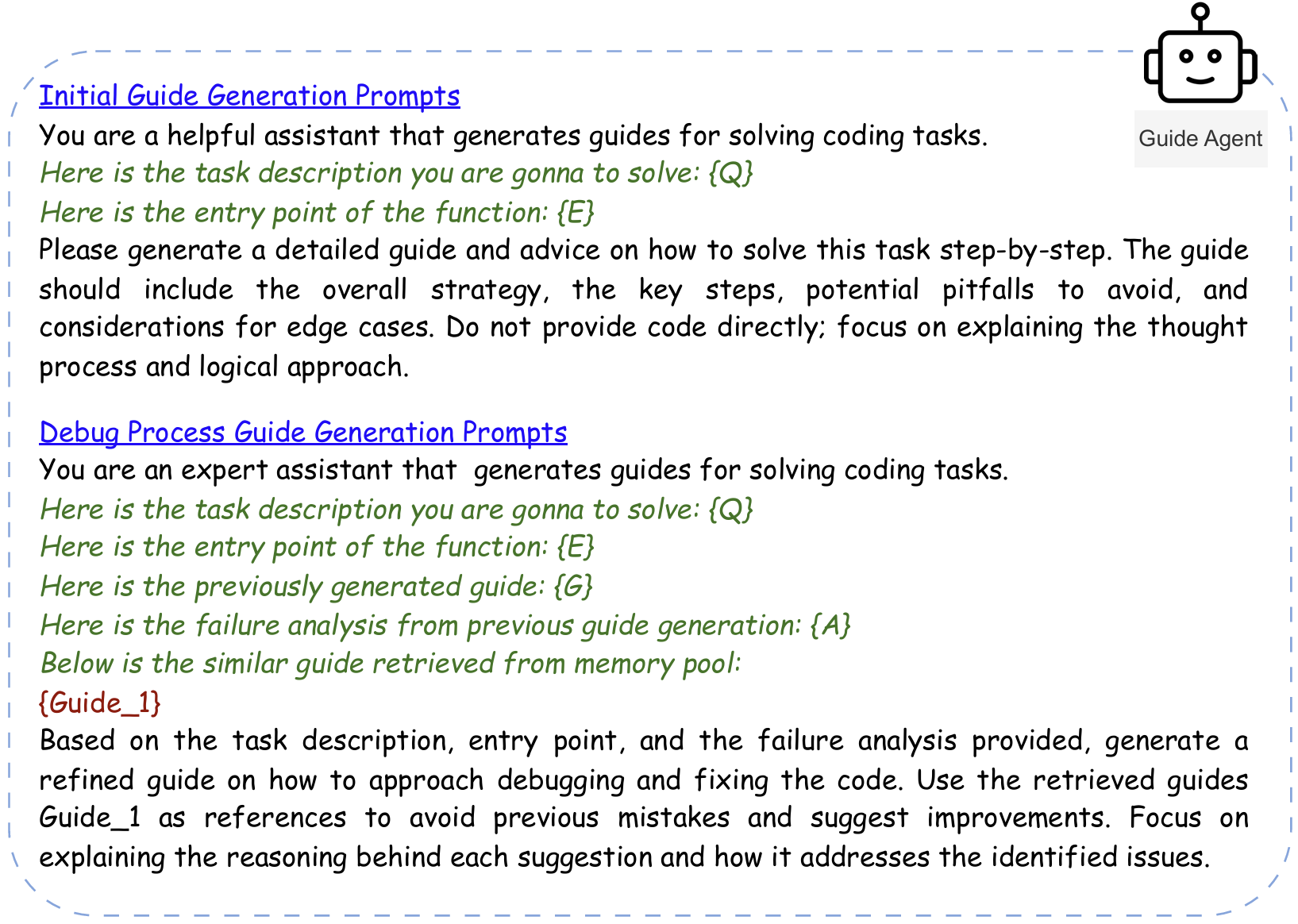} 
    \caption{Guide Agent Prompt} 
    \label{fig:guide}
\end{figure}
\subsection{Guide LLM Agent}
The Guide Agent is assigned a special role with the primary task of conducting initial reasoning and thinking. It dynamically adjusts the input information based on the task's complexity and the current stage of execution. The Guide LLM is also responsible for selecting and applying relevant guides extracted from the Memory Pool. However, to avoid unnecessary token usage and excessive context that may lead to LLM overhead, the memory pool matching and extraction are not performed during the initial generation. As shown in Figure \ref{fig:guide}, different prompts are used by the Guide LLM at the initial stage and the debug stage. In the initial stage, the Guide Agent generates a Generation Guide based on the task description $Q$ and entry point $E$ without adding any additional information. During the debug stage, samples are matched from the memory pool based on task similarity. The system's matching mechanism includes the following components:

\begin{itemize}
    \item $Q$: The task description.
    \item $G$: The generated guide by the Guide LLM.
    \item $K$: A set of matching keywords extracted from the task description $Q$ and generated code $C$.
    \item $E$: The execution results, which include both visible test cases $T_v$ and hidden test cases $T_h$.
\end{itemize}
The memory pool, denoted as $\mathcal{M}$, stores tuples in the form of $(Q_i, G_i, K_i)$, where:
\begin{itemize}
    \item $Q_i$ is the task description.
    \item $G_i$ is the corresponding generation guide.
    \item $K_i$ is the set of keywords extracted by the GPT-4o-mini API after the task successfully passes both $T_v$ and $T_h$.
\end{itemize}
When a task is completed successfully (it passes both visible and hidden test cases), we call the GPT-4o-mini API to generate the set of keywords $K_i$ based on the task description $Q$ and the generated code $C$. This process is formulated as:
\begin{equation}
    K_i = \text{ExtractKeywords}(Q, C)
\end{equation}
where $\text{ExtractKeywords}$ is the function to extract relevant keywords from the task description and its corresponding solution. When processing a new task $Q_{\text{new}}$ during the Debug phase, the memory pool is queried for similar tasks using an SBERT \cite{reimers2019sentencebertsentenceembeddingsusing} \cite{feng2020codebertpretrainedmodelprogramming} for description match and BM25 \cite{lin2021briefnotesdeepimpactcoil} for the index term match, finally achieve hybrid retrieval systems \cite{karpukhin2020densepassageretrievalopendomain}
. The overall process shown in Algorithm~\ref{alg:memory_pool}, which is the pseudo-code illustrates the memory pool matching functions, where the function $\text{Sim}$ computes the similarity between $Q_{\text{new}}$ and $Q_i$, as well as the keywords.

The Guide LLM generates a final guide $G_{\text{final}}$ by augmenting the initial guide $G_{\text{init}}$ generated from $Q_{\text{new}}$ with the relevant guide $G_i$ retrieved from the memory pool based on the similarity score:
\begin{equation}
    G_{\text{final}} = \text{GuideLLM}({G_{\text{init}}, G_i, Q, A})
\end{equation}
where $A$ represents the failure analysis generated by the Feedback Agent which will be introduced in (Section.~\ref{sec:debug}).
\begin{algorithm}
\caption{Memory Pool Extraction}
\label{alg:memory_pool}
\scriptsize
\begin{algorithmic}[1]
    \Require Memory pool $\mathcal{M} \gets \emptyset$, Guide LLM model, similarity threshold list $\mathcal{T} \gets []$

    \Function{RetrieveFromMemoryPool}{$\mathcal{M}, Q'$}
        \State \textbf{Input:} Memory pool $\mathcal{M}$, New task description $Q'$
        \State $\mathcal{S} \gets []$ {(Initialize similar guides list)}
        \ForAll{$(Q, G, K) \in \mathcal{M}$}
            \State $\tau \gets \text{Sim}(Q', Q, K)$ {(Compute similarity)}
            \If{$|\mathcal{S}| < 3$} 
                \State $\mathcal{S} \gets \mathcal{S} \cup \{(Q, G)\}$
                \State $\mathcal{T} \gets \mathcal{T} \cup \{\tau\}$
            \Else
                \State $\tau_{\min} \gets \min(\mathcal{T})$ 
                \If{$\tau > \tau_{\min}$}
                    \State Remove element with $\tau_{\min}$ from $\mathcal{S}$ and $\mathcal{T}$
                    \State $\mathcal{S} \gets \mathcal{S} \cup \{(Q, G)\}$
                    \State $\mathcal{T} \gets \mathcal{T} \cup \{\tau\}$
                \EndIf
            \EndIf
        \EndFor
        \State \textbf{Return} top 3 guides $\mathcal{S}$ sorted by $S$ in descending order
    \EndFunction
\end{algorithmic}
\normalsize
\end{algorithm}
The use of a memory pool to augment the Guide LLM with relevant past experiences contributes to enhanced contextual understanding. By retrieving and incorporating previously successful guides and related task information, the system can generate more accurate and context-aware guides, which leads to better quality in code generation. This enrichment ensures that the LLM is not starting from scratch for every task but instead builds upon a foundation of prior experience. 


\subsection{Debug LLM Agent}\label{sec:debug}
For the Debug Agent, after receiving the generated Generation Guide, it works in conjunction with the task description $Q$ and the entry point $E$ to facilitate code generation. The prompts used for the initial code generation and for fixing code are different, and during the initial generation phase, there is no matching with the memory pool or any failure analysis involved. This choice is similar to the approach in previous work \cite{zhong2024ldb}, where seed code is generated before debugging to prioritize simpler tasks. However, if require running an initial execution to generate seed code for 500 samples before proceeding with debugging like MBPP dataset—this seed process can be extremely time-consuming. Especially in the case of MBPP-ET, where the length of the test set is several times longer than the original, it can lead to a significant waste of time on execution and computational costs, followed by the manual running of debugging procedures.
\begin{figure}[htbp]
\scriptsize
    \centering
    \includegraphics[width=.4\textwidth]{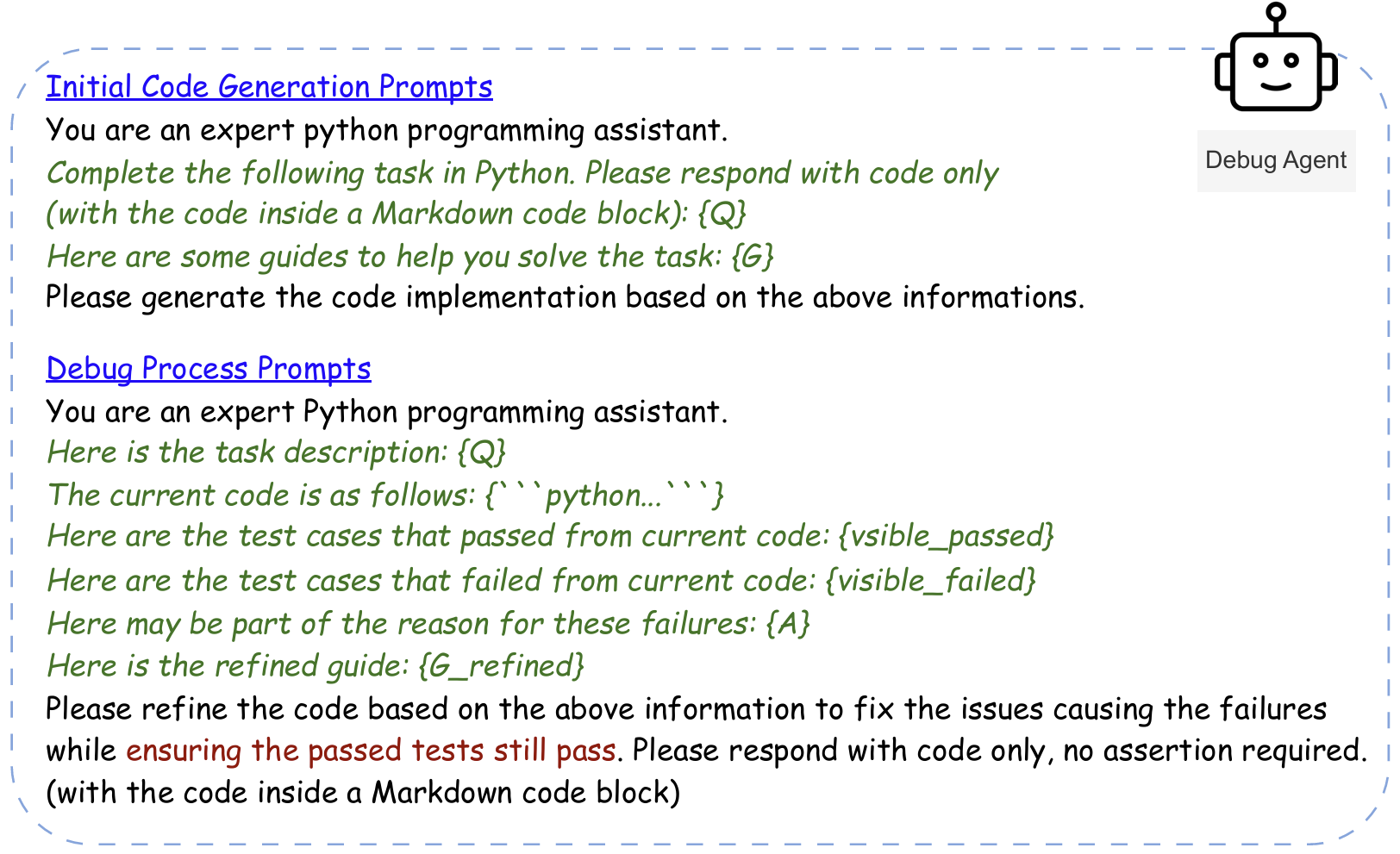} 
    \caption{Debug Agent Prompt} 
    \label{fig:debug}
\end{figure}
Conversely, for smaller dataset HumanEval, this approach can accelerate the overall debugging process. In RGD, simpler tasks are filtered out in the first round, thereby streamlining the debugging process. Figure \ref{fig:debug} illustrates the corresponding prompts used at different stages. By efficiently identifying and eliminating simpler tasks early, the RGD framework ensures that computational resources are focused on more complex problems, enhancing both the speed and effectiveness of the debugging process.
\begin{figure}[htbp]
\scriptsize
    \centering
    \includegraphics[width=.4\textwidth]{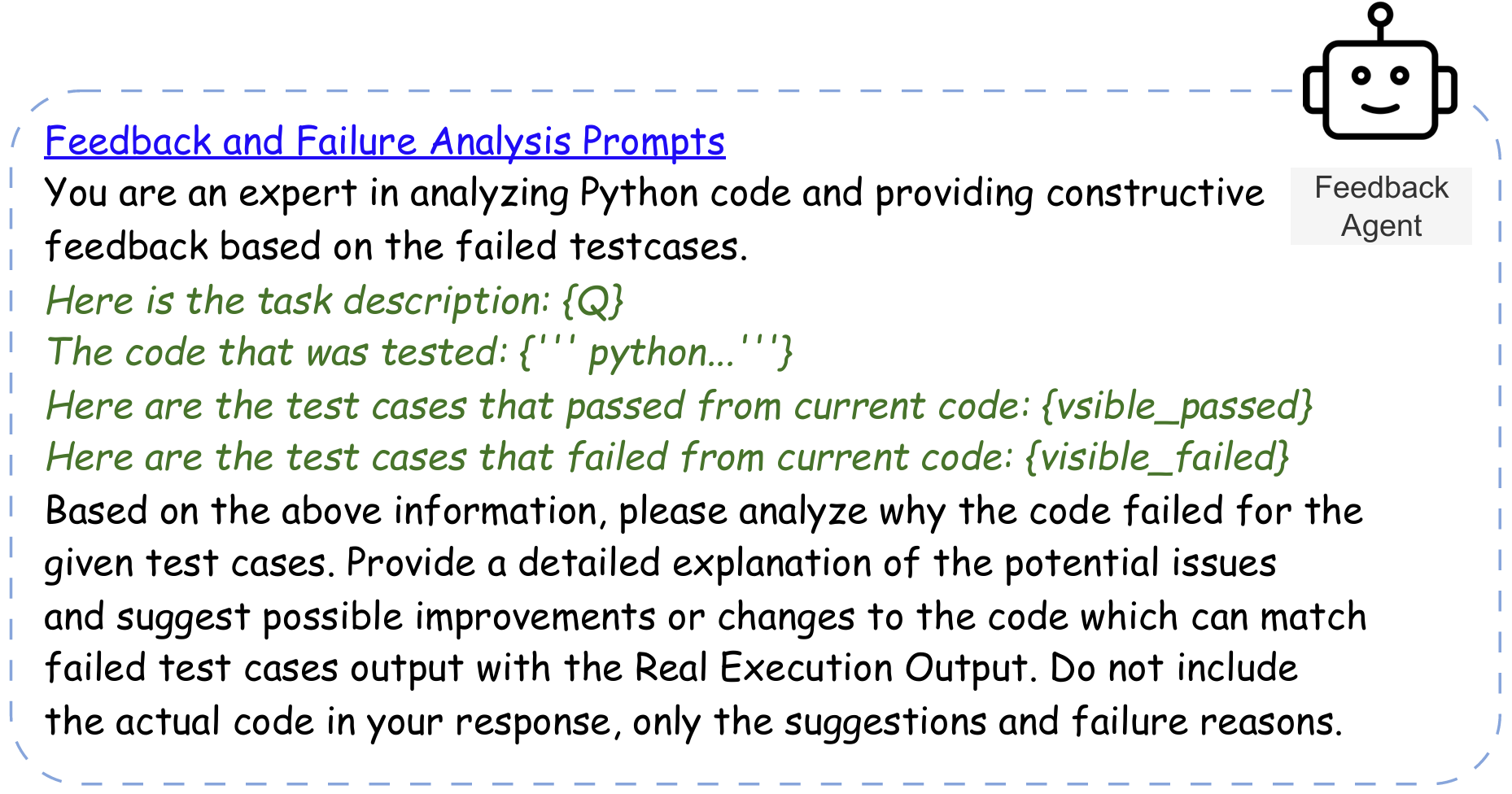} 
    \caption{Feedback Agent Prompt} 
    \label{fig:feedback}
\end{figure}
\subsection{Feedback LLM Agent}
The Feedback Agent is quite similar to the previous reflection strategy \cite{shinn2023reflexionlanguageagentsverbal}, where the LLM analyzes the code errors based on execution results to achieve self-repair. In the RGD architecture, the Feedback Agent plays a similar role by first recording both the failed and successful test cases and then consolidating them based on the execution results. If an exception occurs during execution, the type of exception is also recorded as part of the Real Execution Output. Subsequently, both the passed visible test cases and the failed visible test cases are provided to the Feedback LLM for analysis. This analysis also includes the failed code and the task description, as shown in Figure \ref{fig:feedback}. The final feedback generates a failure analysis, which serves as auxiliary information for both the Guide LLM Agent and the Debug LLM Agent.

Directly regenerating code based on execution results and failed code often results in output that is far inferior to handling a single task \cite{moon2024coffeeboostcodellms}. The Feedback Agent is responsible for independently processing the failure analysis and providing possible modification suggestions. By isolating the failure analysis and proposed corrections, the Feedback Agent enables the LLMs to perform better when dealing with complex tasks that require nuanced debugging and iteration.
\section{Experiment Setup\label{cha:setup}}
\definecolor{JapaneseLaurel}{rgb}{0,0.498,0}
\definecolor{JapaneseLaurel1}{rgb}{0,0.501,0}
\begin{table*}
\scriptsize
\centering
\begin{tblr}{
  row{1} = {c},
  row{4} = {c},
  row{5} = {c},
  row{6} = {c},
  row{7} = {c},
  row{8} = {c},
  row{9} = {c},
  row{10} = {c},
  row{11} = {c},
  row{12} = {c},
  row{13} = {c},
  row{14} = {c},
  row{15} = {c},
  cell{1}{1} = {r=3}{},
  cell{1}{2} = {r=3}{},
  cell{1}{3} = {c=16}{},
  cell{2}{4} = {c=2}{c},
  cell{2}{6} = {c},
  cell{2}{7} = {c=2}{c},
  cell{2}{9} = {c},
  cell{2}{10} = {c=2}{c},
  cell{2}{12} = {c},
  cell{2}{13} = {c=2}{c},
  cell{2}{15} = {c},
  cell{2}{16} = {c=2}{c},
  cell{2}{18} = {c},
  cell{3}{4} = {c},
  cell{3}{5} = {c},
  cell{3}{6} = {c},
  cell{3}{7} = {c},
  cell{3}{8} = {c},
  cell{3}{9} = {c},
  cell{3}{10} = {c},
  cell{3}{11} = {c},
  cell{3}{12} = {c},
  cell{3}{13} = {c},
  cell{3}{14} = {c},
  cell{3}{15} = {c},
  cell{3}{16} = {c},
  cell{3}{17} = {c},
  cell{3}{18} = {c},
  cell{4}{1} = {r=6}{},
  cell{5}{5} = {fg=JapaneseLaurel},
  cell{5}{8} = {fg=red},
  cell{5}{11} = {fg=JapaneseLaurel},
  cell{5}{14} = {fg=JapaneseLaurel1},
  cell{5}{17} = {fg=JapaneseLaurel},
  cell{6}{5} = {fg=JapaneseLaurel},
  cell{6}{8} = {fg=JapaneseLaurel},
  cell{6}{11} = {fg=red},
  cell{6}{14} = {fg=JapaneseLaurel1},
  cell{6}{17} = {fg=JapaneseLaurel},
  cell{7}{5} = {fg=JapaneseLaurel},
  cell{7}{8} = {fg=JapaneseLaurel},
  cell{7}{11} = {fg=JapaneseLaurel1},
  cell{7}{14} = {fg=JapaneseLaurel1},
  cell{7}{17} = {fg=JapaneseLaurel},
  cell{8}{5} = {fg=JapaneseLaurel},
  cell{8}{8} = {fg=JapaneseLaurel},
  cell{8}{11} = {fg=JapaneseLaurel},
  cell{8}{14} = {fg=JapaneseLaurel1},
  cell{9}{5} = {fg=JapaneseLaurel},
  cell{9}{8} = {fg=JapaneseLaurel},
  cell{9}{11} = {fg=JapaneseLaurel},
  cell{9}{14} = {fg=JapaneseLaurel1},
  cell{9}{17} = {fg=JapaneseLaurel},
  cell{10}{1} = {r=6}{},
  cell{11}{5} = {fg=JapaneseLaurel},
  cell{11}{8} = {fg=JapaneseLaurel},
  cell{11}{11} = {fg=JapaneseLaurel},
  cell{11}{14} = {fg=JapaneseLaurel},
  cell{11}{17} = {fg=red},
  cell{12}{5} = {fg=JapaneseLaurel},
  cell{12}{8} = {fg=JapaneseLaurel},
  cell{12}{11} = {fg=JapaneseLaurel},
  cell{12}{14} = {fg=JapaneseLaurel},
  cell{12}{17} = {fg=JapaneseLaurel},
  cell{13}{5} = {fg=JapaneseLaurel},
  cell{13}{8} = {fg=JapaneseLaurel},
  cell{13}{11} = {fg=JapaneseLaurel},
  cell{13}{14} = {fg=JapaneseLaurel},
  cell{13}{17} = {fg=red},
  cell{14}{5} = {fg=JapaneseLaurel},
  cell{14}{8} = {fg=JapaneseLaurel},
  cell{14}{11} = {fg=JapaneseLaurel},
  cell{14}{14} = {fg=JapaneseLaurel},
  cell{15}{5} = {fg=JapaneseLaurel},
  cell{15}{8} = {fg=JapaneseLaurel},
  cell{15}{11} = {fg=JapaneseLaurel},
  cell{15}{14} = {fg=JapaneseLaurel},
  cell{15}{17} = {fg=JapaneseLaurel},
  hline{1} = {-}{},
  hline{2,4} = {3-18}{},
  hline{3} = {4-5,7-8,10-11,16-17}{0.03em},
  hline{3} = {13-14}{},
  hline{4} = {1-2}{0.03em},
  hline{9,15} = {2-18}{dotted},
  hline{10} = {-}{0.05em},
  hline{16} = {-}{0.08em},
}
\textbf{Model} & \textbf{Approch}        & \textbf{Dataset} &                         &                            &  &                         &                            &  &                         &                            &  &                         &                            &  &                         &                            &  \\
               &                         &                  & \textbf{HumanEval}      &                            &  & \textbf{HumanEval-ET}   &                            &  & \textbf{MBPP}           &                            &  & \textbf{MBPP-ET}        &                            &  & \textbf{APPS-100}       &                            &  \\
               &                         &                  & \textbf{Acc $\uparrow$} & \textbf{$\Delta \uparrow$} &  & \textbf{Acc $\uparrow$} & \textbf{$\Delta \uparrow$} &  & \textbf{Acc $\uparrow$} & \textbf{$\Delta \uparrow$} &  & \textbf{Acc $\uparrow$} & \textbf{$\Delta \uparrow$} &  & \textbf{Acc $\uparrow$} & \textbf{$\Delta \uparrow$} &  \\
GPT-4o         & Direct                  &                  & 87.8                    &                            &  & 84.7                    &                            &  & 67.2                    &                            &  & 56.2                    &                            &  & 56.5                    &                            &  \\
               & COT                     &                  & 92.6                    & +4.8                       &  & 82.3                    & -2.4                       &  & 68.2                    & -1.0                       &  & 58.6                    & +2.5                       &  & 58.5                    & +3.0                       &  \\
               & Self-Planning           &                  & 90.2                    & +2.4                       &  & 90.8                    & +6.1                       &  & 59.3                    & -7.9                       &  & 55.8                    & +0.4                       &  & 58.0                    & +1.5                       &  \\
               & Self-Debugging (+Trace) &                  & 89.0                    & +1.2                       &  & 91.4                    & +6.7                       &  & 66.2                    & +1.0                       &  & 61.0                    & +4.8                       &  & 60.0                    & +3.5                       &  \\
               & LDB                     &                  & 94.5                    & +6.7                       &  & 91.4                    & +6.7                       &  & 74.8                    & +7.6                       &  & 71.0                    & +14.8                      &  & -                       & -                          &  \\
               & \textbf{RGD (ours)}     &                  & \textbf{97.6}           & \textbf{+9.8}              &  & \textbf{97.6}           & \textbf{+12.9}             &  & \textbf{83.4}           & \textbf{+16.2}             &  & \textbf{77.8}           & \textbf{+21.6}                      &  & \textbf{63.0}           & \textbf{+6.5}              &  \\
GPT-4o-mini    & Direct                  &                  & 87.8                    &                            &  & 77.4                    &                            &  & 57.7                    &                            &  & 44.6                    &                            &  & 47.4                    &                            &  \\
               & COT                     &                  & 89.6                    & +1.8                       &  & 78.6                    & +1.2                       &  & 58.2                    & +0.6                       &  & 45.3                    & +0.7                       &  & 45.0                    & -2.4                       &  \\
               & Self-Planning           &                  & 89.0                    & +1.2                       &  & 88.3                    & +10.9                      &  & 58.4                    & +0.7                       &  & 48.2                    & +3.6                       &  & 49.0                    & +1.6                       &  \\
               & Self-Debugging (+Trace) &                  & 88.4                    & +0.6                       &  & 87.8                    & +10.4                      &  & 64.4                    & +6.7                       &  & 57.4                    & +12.8                      &  & 45.0                    & -2.4                       &  \\
               & LDB                     &                  & 90.2                    & +2.4                       &  & 89.6                    & +12.2                      &  & 73.6                    & +15.9                      &  & 65.8                    & +21.2                      &  & -                       & -                          &  \\
               & \textbf{RGD (ours)}     &                  & \textbf{96.9}           & \textbf{+9.1}              &  & \textbf{97.5}           & \textbf{+20.1}             &  & \textbf{80.6}           & \textbf{+22.9}             &  & \textbf{69.7}           & \textbf{+25.1}             &  & \textbf{56.0}           & \textbf{+8.6}              &  
\end{tblr}
\caption{Pass@1 results for six approaches, the $\Delta$ denote the percentage improvement against the Direct baseline approach. COT \cite{wei2022chain}, Self-Planning \cite{10.1145/3672456}, Self-Debugging (+Trace) \cite{chen2023teaching} tested all five benchmarks, where LDB framework tested only on HumanEval/ET and MBPP/ET benchmarks \cite{zhong2024ldb}}
\label{tab:result}
\end{table*}
\subsection{Dataset}
During the execution process, we use the test sets from various benchmarks to verify if the current code contains any bugs. In RGD, We utilized the HumanEval \cite{chen2021evaluating} and MBPP \cite{austin2021programsynthesislargelanguage} datasets, and followed the approach in \cite{zhong2024ldb} for allocating visible and hidden test cases. In HumanEval, the visible tests are extracted from the task description, while the hidden tests are taken from the dataset's own tests. In MBPP, the first test case is used as the visible test case.

Additionally, we use HumanEval-ET and MBPP-ET \cite{dong2023codescore} to address the limitations of test cases from the original datasets. For these datasets, the ratio of visible to hidden test cases is set to 6:4. The last dataset we used is APPS \cite{hendrycksapps2021}, from which we selected 100 samples for testing with difficulty levels categorized as introductory, interview, and competition in the ratio of 5:3:2. Unlike the previous datasets, APPS relies on input-output results for validation, requiring a unique handling approach when processing the APPS dataset.
\subsection{Method}
We compared RGD with five different approaches. Direct is the baseline measurement, where code is generated directly based on the task description. Chain-of-Thought \cite{wei2022chain} enhances the model's self-planning ability by adding step-by-step prompts. Self-Planning \cite{10.1145/3672456} is based on the idea of planning first and then reasoning. Self-Debugging \cite{chen2023teaching} involves self-reflection by incorporating feedback from execution results. Lastly, we also tested the LDB framework \cite{zhong2024ldb}.

We conducted evaluations using the Pass@1 metric, In the set $k$ iterations (10 in our experiment), the problem is considered solved as long as it is solved successfully once. We primarily tested the currently most outperforming GPT-4o model and the GPT-4o-mini model.
\section{Experiment Result and Analysis \label{cha:result}}
\subsection{Main Result}
The main experimental results are presented in Table \ref{tab:result}. Here, $Acc$ represents the Pass@1 accuracy, and $\Delta$ denotes the percentage improvement over the direct code generation approach based on the task description. We can observe that RGD demonstrates state-of-the-art performance across all datasets, with further improvements compared to the LDB approach. For the HumanEval dataset with GPT-4o, RGD shows an improvement of 9.8\% over the baseline and 3.1\% over LDB. Our experiments reveal that RGD achieves particularly significant improvements on the HumanEval-ET and MBPP-ET benchmarks, which contain more test cases, with a maximum improvement of 25.1\% on MBPP-ET. This performance boost is mainly due to the limited number of visible test cases in the original datasets, leading to situations where code that passes the specific tests still contains vulnerabilities and fails the hidden test cases, this highlights the critical role of the Feedback Agent's analysis throughout the debugging process.

For the APPS dataset, we conducted experiments using a total of 100 samples. We did not test the LDB framework on the APPS benchmark because it would require significant modifications to its program. Overall, it is evident that RGD achieves enhancements over other methods, with a maximum improvement of 8.6\% in the GPT-4o-mini model.
\begin{table}
\scriptsize
\centering
\begin{tblr}{
  cells = {c},
  cell{1}{1} = {r=3}{},
  cell{1}{2} = {c=8}{},
  cell{2}{2} = {c=2}{},
  cell{2}{5} = {c=2}{},
  cell{2}{8} = {c=2}{},
  hline{1,4,6} = {-}{},
  hline{2} = {2-9}{},
  hline{3} = {2-3,5-6,8-9}{},
}
Benchmark & Component Remove From RGD &       &  &             &       &  &                  &       \\
          & Memory Pool               &       &  & Guide Agent &       &  & Failure Feedback &       \\
          & Acc                       & Drop  &  & Acc         & Drop  &  & Acc              & Drop  \\
HumanEval & 95.7\%                    & 1.9\% &  & 93.2\%      & 4.4\% &  & 95.1\%           & 2.5\% \\
MBPP      & 78.4\%                    & 5.0\% &  & 77.0\%      & 6.4\% &  & 73.6\%           & 9.8\% 
\end{tblr}
\caption{Ablation study on the RGD framework using the GPT-4o model, evaluated on the HumanEval and MBPP}
\label{tab:ablation}
\end{table}
\subsection{Ablations Studies}
We conducted ablation studies to validate the effectiveness of our approach. We selected GPT-4o for experiments on two benchmarks, HumanEval and MBPP, and compared the results with the original RGD framework's performance on these benchmarks, which were 97.6 pass@1 for HumanEval and 83.4 pass@1 for MBPP, respectively. We performed ablation tests under three different scenarios:

\textbf{Memory Pool Removal}: In this scenario, we removed the memory pool, and the Guide Agent no longer retrieved cases from it to generate and refine guides. As shown in Table \ref{tab:ablation}, this resulted in a drop of 1.9 percentage points on the HumanEval dataset and a drop of 5 percentage points on the MBPP dataset.

\textbf{Guide Agent Removal}: In this case, we removed the entire Guide Agent component, this led to a 4.4\% drop on the HumanEval dataset and a 6.4\% drop on the MBPP dataset. The results clearly demonstrate the significant impact of the Guide Agent on the overall performance of the RGD framework.

\textbf{Failure Feedback Removal}: Here, we removed the Failure Feedback component, and the Feedback LLM did not generate Failure Analysis to assist with debugging and refinement. The results showed performance drops on both datasets, especially on the MBPP dataset, where the decrease was as high as 9.8\%.

The ablation study results confirm that the inclusion of the Memory Pool, Guide Agent, and Failure Feedback significantly enhances the overall performance of the RGD framework. Furthermore, we observe a trend from the drop percentages: datasets with more samples (e.g., MBPP) tend to experience a more substantial impact from the ablations. This also corroborates that the RGD framework maintains excellent performance when dealing with a larger number of tests. Both the Memory Pool and Failure Feedback influence subsequent refinement and debugging, and the information in the memory pool is accumulated during execution. This explains why the improvement on MBPP (500 samples) is greater than on HumanEval (163 samples).

\section{Conclusion \label{cha:conclusion}}
In this paper, we present RGD, a novel framework designed to enhance code generation by leveraging a Memory Pool for refining Generation Guides and incorporating feedback through a dedicated Feedback Agent. Our approach allows LLMs to iteratively refine generated code by utilizing previously stored memory and dynamic failure analysis. By selectively extracting relevant guides from the Memory Pool and continuously refining them based on task similarity, RGD significantly boosts the accuracy of code generation across multiple benchmarks. Experiments demonstrate that RGD achieves state-of-the-art performance in code generation and debugging. We anticipate that our work further demonstrates and enhances the capability of LLMs to learn from past, and efficiently adapt to new challenges.
\bibliographystyle{ieeetr}
\bibliography{ref.bib}

\end{document}